\documentclass[twocolumn, aps, pra, prl, superscriptaddress, preprintnumbers, letterpaper]{revtex4-1}

\usepackage{amsthm}
\usepackage{amsmath}
\usepackage{amssymb}
\usepackage{braket}
\usepackage{bbold}
\usepackage{graphicx}
\usepackage{subfigure}
\usepackage{hyperref}
\usepackage{appendix}
\usepackage{tikz}
\usepgflibrary{shapes.geometric}
\usetikzlibrary{arrows}

\theoremstyle{definition}

\hypersetup{colorlinks=true, citecolor=blue, urlcolor=blue, linkcolor=blue}

\begin{document}
\title{Parameterized Bipartite Entanglement Measure}
\author{Zhi-Wei Wei}
\email{weizhw@cnu.edu.cn}
\affiliation{School of Mathematical Sciences, Capital Normal University, 100048 Beijing, China}
\author{Shao-Ming Fei}
\email{feishm@cnu.edu.cn}
\affiliation{School of Mathematical Sciences, Capital Normal University, 100048 Beijing, China}
\affiliation{Max-Planck-Institute for Mathematics in the Sciences, 04103 Leipzig, Germany}

\bigskip

\begin{abstract}
We propose a novel parameterized entanglement measure $\alpha$-concurrence for bipartite systems. By employing positive partial transposition and realignment criteria, we derive analytical lower bounds for the $\alpha$-concurrence. Moreover, we calculate explicitly the analytic expressions of the $\alpha$-concurrence for isotropic states and Werner states.
\end{abstract}

\maketitle

\section{introduction}
As a distinguishing feature of quantum mechanics, quantum entanglement is an important  resource \cite{schrodinger1935gegenwartige,PhysRev.47.777,PhysPhys195} in quantum computation and quantum information processing \cite{nielsen2002quantum} such as quantum dense coding \cite{PhysRevLett.69.2881}, clock synchronization \cite{PRL852010}, quantum teleportation \cite{PhysRevLett.70.1895}, quantum secret sharing \cite{PhysRevA.59.1829} and quantum cryptography \cite{PRL67661}.
One of the problems in quantum entanglement theory is to quantify the entanglement of a bipartite system. A reasonable entanglement measure $E$ should fulfill
\cite{PhysRevLett.78.2275,PhysRevA.57.1619,PhysRevA.56.R3319,PhysRevA.59.141,PRL842014}:
(E1) $E\left(\rho\right)\geqslant0$, with the equality holding iff $\rho$ is separable;
(E2) $E$ is convex, i.e., $E\left(t\rho+\left(1-t\right)\sigma\right)\leqslant tE\left(\rho\right)+\left(1-t\right)E\left(\sigma\right)$ for any states $\rho$ and $\sigma$, where $t\in[0,1]$;
(E3) $E$ does not increase under local operation and classical communication (LOCC), i.e., $E\left(\rho\right)\geqslant E\left(\Lambda\left(\rho\right)\right)$ for any
completely positive and trace preserving (CPTP) LOCC $\Lambda$;
(E4) $E$ does not increase on average under stochastic LOCC, i.e., $E\left(\rho\right)\geqslant\sum_kp_kE\left(\rho_k\right)$, where $p_k=\mathrm{Tr}A_k\rho A_k^{\dagger}$, $\rho_k=A_k\rho A_k^{\dagger}/p_k$ and $A_k$ are the Kraus operators such that $\sum_kA_k^{\dagger}A_k=I$ with $I$ the identity.

Several reasonable entanglement measures have been presented in the past few years, such as concurrence \cite{PhysRevLett.78.5022,PhysRevA.64.042315}, entanglement of formation \cite{PhysRevA.54.3824,PhysRevLett.80.2245,horodecki2001entanglement},
negativity \cite{PhysRevA.65.032314,PhysRevLett.95.090503} and R\'enyi-$\alpha$ entropy of entanglement \cite{gour2007dual,Kim_2010}. Generally it is formidably difficult to calculate analytically the degree of entanglement for arbitrary given states. Analytical results are often available for two-qubit states \cite{PhysRevLett.80.2245} or special higher-dimensional mixed states \cite{PhysRevLett.85.2625,PhysRevA.64.062307,PhysRevA.67.012307} for several entanglement measures \cite{PhysRevA.68.062304,buchholz2016evaluating}. Efforts have been also made towards the analytical lower bounds of entanglement measures like the concurrence \cite{PhysRevLett.95.210501,Li_Guo_2009}. In \cite{PhysRevLett.95.040504} analytical lower bound of the concurrence has been derived by employing the positive partial transpose (PPT) and realignment criteria \cite{PhysRevLett.77.1413,PhysRevA.59.4206,rudolph2004computable,rudolph2005further,chen2003matrix}.
In this paper, inspired by Tsallis-$q$ entropy of entanglement \cite{raggio1995properties,PhysRevA.81.062328} and parameterized entanglement monotone $q$-concurrence \cite{PhysRevA.103.052423}, we propose a novel parameterized entanglement measure $\alpha$-concurrence for any $0\leqslant\alpha\leqslant1/2$. Based on PPT and realignment criteria, we also obtain an analytic lower bound of the $\alpha$-concurrence for general bipartite systems. Finally, we calculate the $\alpha$-concurrence of isotropic states and Werner states analytically.

\section{$\alpha$-concurrence}\label{ac}
Let $\mathcal{H}_A\otimes\mathcal{H}_B$ be an arbitrary $d\times d$ dimensional bipartite Hilbert space associated with subsystems $A$ and $B$. Any pure state on the $\mathcal{H}_A\otimes\mathcal{H}_B$ can be written as in the Schmidt form,
\begin{equation}\label{ac1}
\ket{\psi}=\sum_{i=1}^r\sqrt{\lambda_i}\ket{a_ib_i},
\end{equation}
where $\sum_{i=1}^r\lambda_i=1$ with $\lambda_i>0$, $r$ is the Schmidt rank, $1\leqslant r\leqslant d$. $\left\{\ket{a_i}\right\}$ and $\left\{\ket{b_i}\right\}$ are the local bases associated with the subsystems $A$ and $B$, respectively \cite{nielsen2002quantum}.

$\mathit{Definition}$. For any pure state $\ket{\psi}$ given in (\ref{ac1}), the $\alpha$-concurrence is defined by
\begin{equation}\label{text11}
C_{\alpha}\left(\ket{\psi}\right)=\mathrm{Tr}\rho_A^{\alpha}-1
\end{equation}
for any $0\leqslant \alpha\leqslant1/2$, where $\rho_A=\mathrm{Tr}_B|\psi\rangle\langle\psi|$.

From (\ref{text11}), for a pure state $\ket{\psi}$ given by (\ref{ac1}) one has
\begin{equation}
C_{\alpha}\left(\ket{\psi}\right)=\sum_{i=1}^r\lambda_i^{\alpha}-1,
\end{equation}
where $C_{\alpha}\left(\ket{\psi}\right)$ satisfies $0\leqslant C_{\alpha}\left(\ket{\psi}\right)\leqslant d^{1-\alpha}-1$. It is obvious that the lower bound is attained if and only if $\ket{\psi}$ is a separable state, that is, $\ket{\psi}=\ket{a_ib_i}$ for some $\ket{a_i}$ and $\ket{b_i}$. While the upper bound is achieved for the maximally entangled pure states $\ket{\Psi^+}=\frac{1}{\sqrt{d}}\sum_{i=1}^d\ket{a_ib_i}$.

For a general mixed state $\rho$ on the Hilbert space $\mathcal{H}_A\otimes\mathcal{H}_B$, the $\alpha$-concurrence is given by the convex-roof extension,
\begin{equation}\label{text14}
C_{\alpha}\left(\rho\right)=\min_{\left\{p_i,\ket{\psi_i}\right\}}
\sum_ip_iC_{\alpha}\left(\ket{\psi_i}\right),
\end{equation}
where the infimum is taken over all possible pure-state decompositions of $\rho=\sum_ip_i|\psi_i\rangle\langle\psi_i|$, with $\sum_ip_i=1$ and $p_i> 0$.
Before showing that $C_{\alpha}\left(\rho\right)$ defined in (\ref{text14}) is indeed a bona fide entanglement measure, we first present the following lemma, see proof in Appendix \ref{a}.

$\mathit{Lemma\ 1}$.
The function $F_{\alpha}\left(\rho\right)=\mathrm{Tr}\rho^{\alpha}-1$ is concave, that is,
\begin{equation}\label{con2}
F_{\alpha}\left(\sum_ip_i\rho_i\right)\geqslant\sum_ip_iF_{\alpha}\left(\rho_i\right)
\end{equation}
for any $0\leqslant \alpha \leqslant1/2$, where $\set{p_i}$ is a probability distribution and $\rho_i$ are density matrices. The equality holds if and only if all $\rho_i$ are the same for all $p_i>0$.

By using the above Lemma 1, we have the following theorem.

$\mathit{Theorem\,1}$.
The $\alpha$-concurrence $C_{\alpha}\left(\rho\right)$ given in (\ref{text14}) is a well defined parameterized entanglement measure.

$\mathit{Proof}$.
We need to verify that $C_{\alpha}\left(\rho\right)$ fulfills the following four requirements.

(E1)
If $\rho$ is an entangled state, then there is at least one entangled pure state $\ket{\psi}$ in any pure state decomposition of $\rho$. Thus $C_{\alpha}\left(\rho\right)>0$. Otherwise, $C_{\alpha}\left(\rho\right)=0$ for separable states.

(E2)
Consider $\rho=t\rho_1+\left(1-t\right)\rho_2$. Let $\rho_1=\sum_ip_i|\psi_i\rangle\langle\psi_i|$ ($\rho_2=\sum_jq_j|\phi_j\rangle\langle\phi_j|$) be the optimal pure state decomposition of $C_{\alpha}\left(\rho_1\right)$ ($C_{\alpha}\left(\rho_2\right)$) with $\sum_ip_i=1$ ($\sum_jq_j=1$) and $p_i>0$ ($q_j>0$). We have
\begin{align}
C_{\alpha}\left(\rho\right)&\leqslant\sum_{i=1}^ktp_iC_{\alpha}\left(\ket{\psi_i}\right)+\sum_{j=1}^l\left(1-t\right)q_jC_{\alpha}\left(\ket{\phi_j}\right)\nonumber\\
&=tC_{\alpha}\left(\rho_1\right)+\left(1-t\right)C_{\alpha}\left(\rho_2\right),
\end{align}
where the first inequality is due to that $\sum_{i=1}^ktp_i|\psi_i\rangle\langle\psi_i|
+\sum_{j=1}^l\left(1-t\right)q_j|\phi_j\rangle\langle\phi_j|$ is also a pure state decomposition of $\rho$.

(E3)
We adopt the approach given in \cite{MINTERT2005207} to show that our entanglement measure does not increase under LOCC. Denote $\vec{\lambda}_{\psi}$ ($\vec{\lambda}_{\phi}$) the Schmidt vector given by the squared Schmidt coefficients of the state $\ket{\psi}$ ($\ket{\phi}$) in the decreasing order. It has been shown that the state $\ket{\phi}$ can be prepared starting from the state $\ket{\psi}$ under LOCC if and only if $\vec{\lambda}_{\psi}$ is majorized by $\vec{\lambda}_{\phi}$ \cite{PhysRevLett.83.436}, $\vec{\lambda}_{\psi}\prec\vec{\lambda}_{\phi}$, where the majorization means that the components $\left[\lambda_{\psi}\right]_i$ ($\left[\lambda_{\phi}\right]_i$) of $\vec{\lambda}_{\psi}$ ($\vec{\lambda}_{\phi}$), listed in nonincreasing order, satisfy $\sum_{i=1}^j\left[\lambda_{\psi}\right]_i\leqslant\sum_{i=1}^j\left[\lambda_{\phi}\right]_i$ for $1< j\leqslant d$, with equality for $j=d$.

Since the entanglement cannot increase under LOCC, any entanglement measure $E$ has to satisfy that $E\left(\psi\right)\geqslant E\left(\phi\right)$ whenever $\vec{\lambda}_{\psi}\prec\vec{\lambda}_{\phi}$. This condition, known as the $\mathit{Schur\, concavity}$, is satisfied if and only if $E$, given as a function of the squared Schmidt coefficients $\lambda_i$'s \cite{ANDO1989163}, is invariant under the permutations of any two arguments and satisfies
\begin{align}\label{condition}
\left(\lambda_i-\lambda_j\right)\left(\frac{\partial E}{\partial\lambda_i}-\frac{\partial E}{\partial\lambda_j}\right)\leqslant0
\end{align}
for any two components $\lambda_i$ and $\lambda_j$ of $\vec{\lambda}$.

For any pure state $\ket{\psi}$ given by (\ref{ac1}), the $\alpha$-concurrence is obviously invariant under the permutations of the Schmidt coefficients for any $0\leqslant\alpha\leqslant 1/2$. Since
\begin{align}
&\left(\lambda_i-\lambda_j\right)\left(\frac{\partial C_{\alpha}}{\partial\lambda_i}-\frac{\partial C_{\alpha}}{\partial\lambda_j}\right)\nonumber\\
&=\alpha\left(\lambda_i-\lambda_j\right)
\left(\lambda_i^{\alpha-1}-\lambda_j^{\alpha-1}\right)\leqslant 0\nonumber
\end{align}
for any two components $\lambda_i$ and $\lambda_j$ of the squared Schmidt coefficients of $\ket{\psi}$, we have $C_{\alpha}\left(\ket{\psi}\right)\geqslant C_{\alpha}\left(\Lambda\ket{\psi}\right)$ for any LOCC $\Lambda$ and $0\leqslant\alpha\leqslant1/2$.

Next let $\rho=\sum_ip_i|\psi_i\rangle\langle\psi_i|$ be the optimal pure state decomposition of $C_{\alpha}\left(\rho\right)$ with $\sum_ip_i=1$ and $p_i>0$. We obtain
\begin{align}
C_{\alpha}\left(\rho\right)&=\sum_ip_iC_{\alpha}\left(\ket{\psi_i}\right)\nonumber\\
&\geqslant\sum_ip_iC_{\alpha}\left(\Lambda\ket{\psi_i}\right)\nonumber\\
&\geqslant C_{\alpha}\left(\Lambda\left(\rho\right)\right)\nonumber
\end{align}
for any $0\leqslant\alpha\leqslant1/2$, where the last inequality is from the definition (\ref{text14}).

(E4)
Let $\rho=\sum_ip_i|\psi_i\rangle\langle\psi_i|$ be the optimal pure state decomposition of $C_{\alpha}\left(\rho\right)$ with $\sum_ip_i=1$ and $p_i>0$. Consider stochastic LOCC protocol given by Kraus operators $A_k$ with $\sum_kA_k^{\dagger}A_k=I$. We have
\begin{align}\label{theorem142}
C_{\alpha}\left(\rho\right)&=\sum_ip_iC_{\alpha}\left(\ket{\psi_i}\right)\nonumber\\
&\geqslant\sum_{i,k}p_ip_{k|i}C_{\alpha}\left(\ket{\psi_i^k}\right)\nonumber\\
&=\sum_{i,k}p_kp_{i|k}C_{\alpha}\left(\ket{\psi_i^k}\right)\nonumber\\
&=\sum_kp_k\left(\sum_ip_{i|k}C_{\alpha}\left(\ket{\psi_i^k}\right)\right)\nonumber\\
&\geqslant\sum_kp_kC_{\alpha}\left(\rho^k\right),
\end{align}
where $p_{k|i}=\mathrm{Tr}A_k|\psi_i\rangle\langle\psi_i|A_k^{\dagger}$ is the probability of obtaining the outcome $k$ with $\ket{\psi_i^k}=A_k\ket{\psi_i}/\sqrt{p_{k|i}}$, and $p_k=\mathrm{Tr}A_k\rho A_k^{\dagger}$ is the probability of obtaining the outcome $k$ with $\rho^k=A_k\rho A_k^{\dagger}/p_k$. The first inequality is due to the concavity of the Lemma 1, since $\mathrm{Tr}_B|\psi_i\rangle\langle\psi_i|=\sum_kp_{k|i}\mathrm{Tr}_B|
\psi_i^k\rangle\langle\psi_i^k|$. The last inequality is from the definition of (\ref{text14}), since $\sum_ip_{i|k}|\psi_i^k\rangle\langle\psi_i^k|=\rho^k$. $\hfill\qedsymbol$

In \cite{PhysRevA.103.052423} the parameterized entanglement monotone $q$-concurrence $C_q\left(\ket{\psi}\right)$ for any pure state $\ket{\psi}$ defined in (\ref{ac1}) has been introduced,
$C_q\left(\ket{\psi}\right)=1-\mathrm{Tr}\rho_A^q$, where $q\geqslant 2$.
It seems that our $\alpha$-concurrence defined by (\ref{text11}),
$C_{\alpha}\left(\ket{\psi}\right)=\mathrm{Tr}\rho_A^{\alpha}-1$,
is in some sense dual to the $q$-concurrence as the parameter $\alpha\in [0,1/2]$, while the parameter $q\geqslant 2$. Nevertheless, these two concurrences characterize the quantum entanglement in different aspects, even though they are both derived from the Tsallis-$q$ of entanglement \cite{PhysRevA.81.062328}. For large enough $q$, the $q$-concurrence $C_q(\rho)$ converges to the constant $1$ for any entangled state $\rho$, while the $\alpha$-concurrence $C_{\alpha}(\rho)$ not for any $\alpha\in[0,1/2]$. Particularly, for $\alpha=0$, the measure $C_0\left(\ket{\psi}\right)=r-1$ for any pure state $\ket{\psi}$ given in (\ref{ac1}), where $r$ is the Schmidt rank of the state $\ket{\psi}$, which is solely determined by the Schmidt rank of the bipartite pure state $\ket{\psi}$. Therefore, the $\alpha$-concurrences with different $\alpha$ provide different characterizations of the feature of entanglement.

\section{bounds on $\alpha$-concurrence}\label{bound}
Owing to the optimization in the calculation of the entanglement measures, it is generally difficult to obtain analytical expressions of the  entanglement measures for general mixed states. In this section, we derive analytical lower bounds for the $\alpha$-concurrence based on PPT and realignment criteria \cite{PhysRevLett.77.1413,HORODECKI19961,PhysRevA.59.4206,
rudolph2004computable,rudolph2005further,chen2003matrix}.

A bipartite state can be written as $\rho=\sum_{ijkl}\rho_{ij,kl}|ij\rangle\langle kl|$, where the subscripts $i$ and $k$ are the row and column indices for the subsystem $A$, respectively, while $j$ and $l$ are such indices for the subsystem $B$. The PPT criterion says that if the state $\rho$ is separable, then the partial transposed matrix $\rho^{\Gamma}=\sum_{ijkl}\rho_{ij,kl}|il\rangle\langle kj|$ with respect to the subsystem $B$ is non-negative, $\rho^{\Gamma}\geqslant 0$.
While the realignment criterion says that the realigned matrix of $\rho$, $\mathcal{R}\left(\rho\right)=\sum_{ijkl}\rho_{ij,kl}|ik\rangle\langle jl|$, satisfies that $\|\mathcal{R}\left(\rho\right)\|_1\leqslant 1$ if $\rho$ is separable, where $\|X\|_1$ denotes the trace norm of matrix $X$, $\|X\|_1=\mathrm{Tr}\sqrt{XX^{\dagger}}$.

For a pure state $\ket{\psi}$ given by (\ref{ac1}), it is straightforward to obtain that \cite{PhysRevLett.95.040504}
\begin{equation}\label{text1}
1\leqslant\left\|\rho^{\Gamma}\right\|_1
=\left\|\mathcal{R}\left(\rho\right)\right\|_1
=\left(\sum_{i=1}^r\sqrt{\lambda_i}\right)^2\leqslant r,
\end{equation}
where $\rho=|\psi\rangle\langle\psi|$.
In particular, for $\alpha=1/2$, the $1/2$-concurrence becomes $C_{\frac{1}{2}}\left(\ket{\psi}\right)=\sum_{i=1}^r\sqrt{\lambda_i}-1$. One has then
\begin{align}\label{cren3}
C_{\frac{1}{2}}\left(\ket{\psi}\right)\geqslant\frac{\left\|\rho^{\Gamma}\right\|_1-1}{\sqrt{r}+1}
\end{align}
for any pure state $\ket{\psi}$ on the $\mathcal{H}_A\otimes\mathcal{H}_B$.

$\mathit{Theorem\ 2}$.
For any mixed state $\rho$ on the $\mathcal{H}_A\otimes\mathcal{H}_B$, the $\alpha$-concurrence $C_{\alpha}\left(\rho\right)$ satisfies
\begin{equation}\label{text13}
C_{\alpha}\left(\rho\right)\geqslant \frac{d^{1-\alpha}-1}{d-1}\left[\max\left(\|\rho^{\Gamma}\|_1,\|\mathcal{R}
\left(\rho\right)\|_1\right)-1\right].
\end{equation}

$\mathit{ Proof}$.
For a pure state $\ket{\psi}$ given in (\ref{ac1}), let us analyze the monotonicity of the following function,
\begin{equation}
g\left(\alpha\right)=\frac{\sum_{i=1}^r\lambda_i^{\alpha}-1}{r^{1-\alpha}-1}
\end{equation}
for any $0\leqslant\alpha\leqslant1/2$, where $r>1$.
The first derivative of $g\left(\alpha\right)$ with respect to $\alpha$ is given by
\begin{align}\label{text020}
\frac{\partial g}{\partial\alpha}&=\frac{G_{r\alpha}}{\left(r^{1-\alpha}-1\right)^2},
\end{align}
where
\begin{equation}\label{bthem1}
G_{r\alpha}=\sum_i\lambda_i^{\alpha}\ln\lambda_i\left(r^{1-\alpha}-1\right)
+\left(\sum_i\lambda_i^{\alpha}-1\right)r^{1-\alpha}\ln r.
\end{equation}

Employing the Lagrange multiplies \cite{PhysRevA.67.012307} under constraints $\sum_{i=1}^r\lambda_i=1$ and $\lambda_i>0$, one has that there is only one stable point $\lambda_i=1/r$ for every $i=1,\cdots,r$, for which $G_{r\alpha}=0$ for any $0\leqslant\alpha\leqslant1/2$. Since the second derivative at this point,
\begin{align}\label{bthem2}
&\frac{\partial^2G_{r\alpha}}{\partial\lambda_i^2}\Big|_{\lambda_i=\frac{1}{r}}\nonumber\\
&=r^{2-\alpha}\left\{\alpha\left(\alpha-1\right)\ln r+\left(2\alpha-1\right)\left(r^{1-\alpha}-1\right)\right\}\nonumber\\
&<0
\end{align}
for any $0\leqslant\alpha\leqslant1/2$, the maximum extreme value point is just the maximum value point. $g\left(\alpha\right)$ is a decreasing function for any $0\leqslant\alpha\leqslant1/2$, since $\partial g/\partial\alpha\leqslant0$.

We have
\begin{align}\label{text116}
C_{\alpha}\left(\ket{\psi}\right)&\geqslant\frac{r^{1-\alpha}-1}{\sqrt{r}-1}C_{\frac{1}{2}}\left(\ket{\psi}\right)\nonumber\\
&\geqslant \frac{r^{1-\alpha}-1}{r-1}\left(\|\sigma^{\Gamma}\|_1-1\right)\nonumber\\
&\geqslant \frac{d^{1-\alpha}-1}{d-1}\left(\|\sigma^{\Gamma}\|_1-1\right),
\end{align}
where $\sigma=|\psi\rangle\langle\psi|$, the second inequality is due to (\ref{cren3}), the last inequality is due to that $\frac{r^{1-\alpha}-1}{r-1}$ is a decreasing function with respect to $r$.
Assume $\rho=\sum_ip_i|\varphi_i\rangle\langle\varphi_i|$ is the optimal pure state decomposition for $ C_{\alpha}\left(\rho\right)$. Then
\begin{align}\label{text16}
C_{\alpha}\left(\rho\right)&=\sum_ip_iC_{\alpha}\left(\ket{\varphi_i}\right)\nonumber\\
&\geqslant\frac{d^{1-\alpha}-1}{d-1}\sum_ip_i\left(\|\sigma_i^{\Gamma}\|_1-1\right)\nonumber\\
&\geqslant\frac{d^{1-\alpha}-1}{d-1}\left(\|\rho^{\Gamma}\|_1-1\right),
\end{align}
where $\sigma_i=|\varphi_i\rangle\langle\varphi_i|$, the last inequality is due to the convex property of the trace norm and $\|\rho^\Gamma\| \geq 1$ in (\ref{text1}).

Similar to (\ref{text116}) and (\ref{text16}), we obtain from (\ref{text1}) that
\begin{equation}\label{text17}
C_{\alpha}\left(\rho\right)\geqslant\frac{d^{1-\alpha}-1}{d-1}
\left(\|\mathcal{R}\left(\rho\right)\|_1-1\right)
\end{equation}
for any $0\leqslant \alpha \leqslant 1/2$.

Combining (\ref{text16}) and (\ref{text17}), we complete the proof. $\hfill\qedsymbol$

\section{$\alpha$-concurrence for Isotropic and Werner states}\label{acis}
In this section, we compute the $\alpha$-concurrence for isotropic states and Werner states. Let $E$ be a convex-roof extended quantum entanglement measure. Denote $S$ the set of states and $P$ the set all pure states in $S$. Let $G$ be a compact group acting on $S$ by $\left(U,\rho\right)\mapsto U\rho U^{\dagger}$. Assume that the measure $E$ defined on $P$ is invariant under the operations of $G$. One can define the projection $\mathbf{P}$:$ S\rightarrow S$ by $\mathbf{P}\rho=\int dUU\rho U^{\dagger}$ with the standard (normalized) Haar measure $dU$ on $G$, and the function $\eta$ on $\mathbf{P}S$ by
\begin{equation}\label{p30}
\eta\left(\rho\right)=\min\left\{E\left(\ket{\Psi}\right):\ket{\psi}\in P,\,\mathbf{P}|\psi\rangle\langle\psi|=\rho\right\}.
\end{equation}
Then for $\rho\in\mathbf{P}S$, we have
\begin{equation}\label{p31}
E\left(\rho\right)=co\left(\eta\left(\rho\right)\right),
\end{equation}
where $co\left(f\right)$ is the convex-roof extension of a function $f$. In other words, it is the convex hull of $f$.

\subsection{Isotropic states}
The isotropic states $\rho_F$ are given by \cite{PhysRevA.59.4206},
\begin{equation}
\rho_F=\frac{1-F}{d^2-1}\left(I-|\Psi\rangle\langle\Psi|\right)+F|\Psi\rangle\langle\Psi|,
\end{equation}
where $\ket{\Psi}=\frac{1}{\sqrt{d}}\sum_{i=1}^d\ket{ii}$ and $F=\braket{\Psi|\rho_F|\Psi}$. $\rho_F$ is separable if and only if $0\leqslant F\leqslant 1/d$ \cite{PhysRevLett.85.2625}. Inspired by the techniques adopted in \cite{PhysRevLett.85.2625,PhysRevA.67.012307,PhysRevA.64.062307,PhysRevA.68.062304}, we have,
see Appendix \ref{b},
\begin{equation}\label{text32}
\eta_{\alpha}\left(\rho_F\right)=\left(Fd\right)^{1-\alpha}-1
\end{equation}
for any $0\leqslant \alpha\leqslant 1/2$, where $F>1/d$.

Obviously, the second derivative of (\ref{text32}) with respect to $F$ is non-positive for any $0\leqslant\alpha\leqslant1/2$. Hence, $\eta_{\alpha}\left(\rho_F\right)$ is concave in the whole regime $F\in\left(1/d,1\right]$. The $ C_{\alpha}\left(\rho_F\right)$ is the largest convex function that is upper bounded by $\eta_{\alpha}\left(\rho_F\right)$, which is constructed in the following way. Find the line that passes through the points $\left(F=1/d,\eta_{\alpha}=0\right)$ and $\left(F=1,\eta_{\alpha}=d^{1-\alpha}-1\right)$ of $\eta_{\alpha}\left(\rho_F\right)$ for any $0\leqslant\alpha\leqslant1/2$. Thus, we have the following analytical formula of the $\alpha$-concurrence for isotropic states.

$\mathit{Lemma\ 2}$.
The $\alpha$-concurrence for isotropic states $\rho_F\in\mathbb{C}^d\otimes\mathbb{C}^d\ \left(d\geqslant 2\right)$ is given by
\begin{equation}\label{isotropicf1}
C_{\alpha}\left(\rho_F\right)=\begin{cases}
0,       & F\leqslant 1/d,\\[2mm]
\displaystyle\frac{d^{1-\alpha}-1}{d-1}\left(dF-1\right),  &F>1/d,
\end{cases}
\end{equation}
where  $0\leqslant \alpha\leqslant1/2$ and $d\geqslant 2$.

Since $\|\rho_F^{\Gamma}\|_1=\|\mathcal{R}\left(\rho_F\right)\|_1=Fd$ for $F>1/d$ \cite{rudolph2005further, PhysRevA.65.032314}, surprisingly the lower bound of (\ref{text13}) is just exactly the (\ref{isotropicf1}) for every $0\leqslant\alpha\leqslant1/2$ with $d\geqslant2$.

\begin{figure}[t]
   \centering
   \includegraphics[width=8cm]{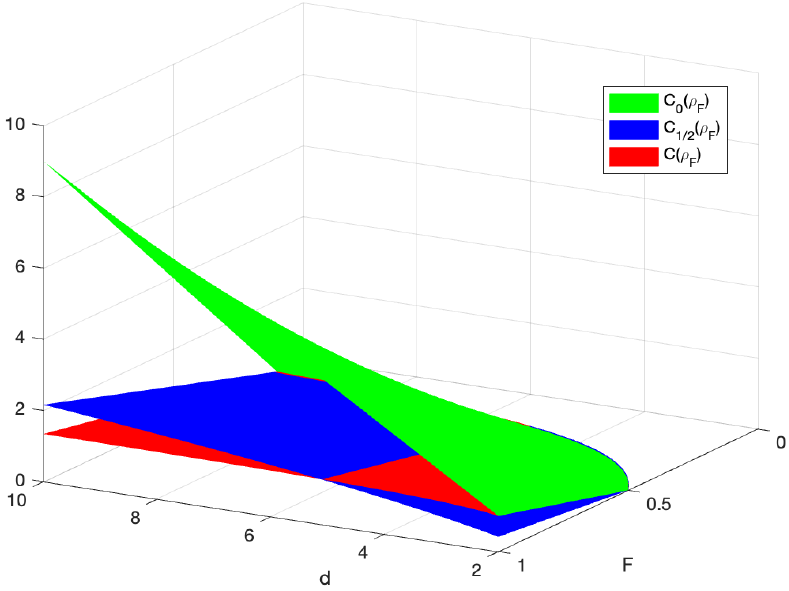}\\
   \caption{Entanglement of isotropic states. The green (blue) surface stands for $C_0\left(\rho_F\right)$ ($C_{1/2}\left(\rho_F\right)$), the red surface is for the concurrence $C\left(\rho_F\right)$.}
  \label{f1}
\end{figure}

The concurrence $C\left(\rho_F\right)$ of isotropic states has been derived in \cite{PhysRevA.67.012307}, $C\left(\rho_F\right)=\sqrt{\frac{2}{d\left(d-1\right)}}\left(dF-1\right)$ for any $F>1/d$. Fig. \ref{f1} exhibits the relations between the concurrence and the $\alpha$-concurrence of isotropic states for $\alpha=0$ and $1/2$.
Especially, it shows that the $C_0\left(\rho_F\right)=C\left(\rho_F\right)$ with $d=2$, and the concurrence of isotropic states is less than the $0$-concurrence with $d>2$. Moreover, we notice that the $1/2$-concurrence is bigger than the concurrence with $d\geq5.1508$.

\subsection{Werner states}
The Werner states are of the form,
\begin{align}\label{p4wer1}
\rho_W=&\frac{2\left(1-W\right)}{d\left(d+1\right)}\left(\sum_{k=1}^d|kk\rangle\langle kk|+\sum_{i<j}|\Psi_{ij}^+\rangle\langle\Psi_{ij}^+|\right)\nonumber\\
&+\frac{2W}{d\left(d-1\right)}\sum_{i<j}|\Psi_{ij}^-\rangle\langle\Psi_{ij}^-|,
\end{align}
where $\ket{\Psi_{ij}^{\pm}}=(\ket{ij}\pm\ket{ji})/\sqrt{2}$ and $W=\mathrm{Tr}(\rho_W\sum_{i<j}|\Psi_{ij}^-\rangle\langle\Psi_{ij}^-|)$ \cite{PhysRevA.68.062304}. $\rho_W$ is separable if and only if $0\leqslant W\leqslant 1/2$ \cite{PhysRevA.64.062307,PhysRevA.40.4277}. For $W>1/2$, we have, see Appendix \ref{c},
\begin{equation}\label{text32wer}
\eta_{\alpha}\left(\rho_W\right)=\left(2W\right)^{1-\alpha}-1
\end{equation}
for any $0\leqslant \alpha\leqslant 1/2$.

It is direct to verify that the second derivative of (\ref{text32wer}) with respect to $W$ is non-positive, namely, $\eta_{\alpha}\left(\rho_W\right)$ is concave. Similar to (\ref{isotropicf1}), we have

$\mathit{Lemma\ 3}$.
The $\alpha$-concurrence for Werner states $\rho_W\in\mathbb{C}^d\otimes\mathbb{C}^d\ \left(d\geqslant 2\right)$ is given by
\begin{equation}\label{wernerf1}
C_{\alpha}\left(\rho_W\right)=\begin{cases}
0,       & W\leqslant 1/2,\\[1mm]
\left(2^{1-\alpha}-1\right)\left(2W-1\right),  &W>1/2,
\end{cases}
\end{equation}
where  $0\leqslant \alpha\leqslant1/2$.

We remark that for $W>1/2$, the lower bound of (\ref{text13}) for Werner states is given by
\begin{align}\label{wernerf2}
C_{\alpha}\left(\rho_W\right)\geqslant\frac{2\left(d^{1-\alpha}-1\right)}
{d\left(d-1\right)}\left(2W-1\right).
\end{align}
Accounting to (\ref{wernerf1}), we obtain
\begin{align}\label{wernerf3}
\left(2^{1-\alpha}-1\right)\left(2W-1\right)\geqslant\frac{2\left(d^{1-\alpha}-1\right)}{d\left(d-1\right)}\left(2W-1\right),
\end{align}
where equality holds if $d=2$, and the inequality holds strictly for higher dimensional quantum systems.

\begin{figure}[t]
   \centering
   \includegraphics[width=8cm]{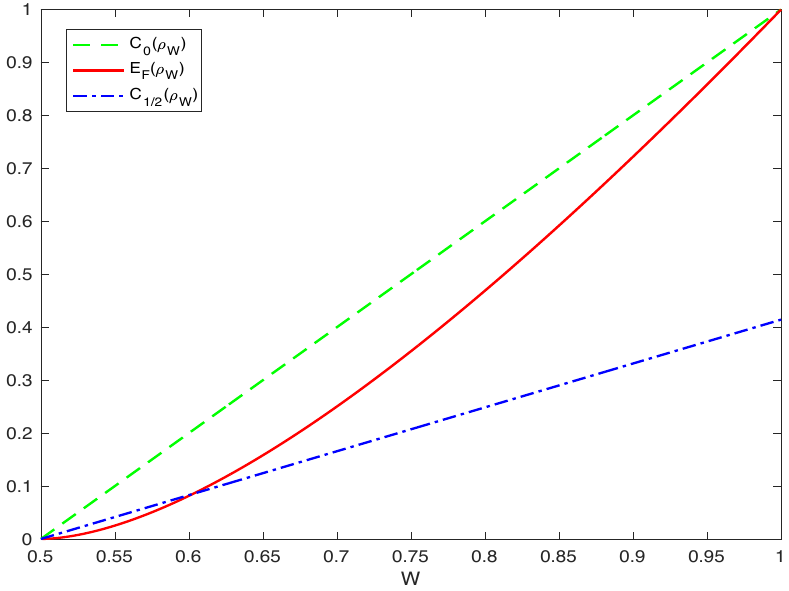}\\
   \caption{Entanglement for Werner states. The dashed (green) line stands for $C_0\left(\rho_W\right)=C\left(\rho_W\right)$. The solid (red) line is for the entanglement of formation $E_F\left(\rho_W\right)$, the dot dashed (blue) line is for the $1/2$-concurrence $C_{1/2}\left(\rho_W\right)$.}
  \label{f2}
\end{figure}

The concurrence of Werner states has been obtained in \cite{PhysRevA62044302}, $C\left(\rho_W\right)=2W-1$ for $W>1/2$. It is direct to find that  $C_{\alpha}\left(\rho_W\right)=\left(2^{1-\alpha}-1\right)C\left(\rho_W\right)$ for any $0\leq\alpha\leq1/2$. Moreover, the entanglement of formation for Werner states is given by \cite{PhysRevA.64.062307}, $E_F\left(\rho_W\right)=H_2[\frac{1}{2}(1-2\sqrt{W\left(1-W\right)})]$.
In Fig. \ref{f2} we illustrate the relations among the concurrence, entanglement of formation and $\alpha$-concurrence of Werner states for $\alpha=0$ and $1/2$.
The entanglement of formation for Werner states $E_F\left(\rho_W\right)$ is always upper bounded by the $C_0\left(\rho_W\right)=C\left(\rho_W\right)$, and larger than the $1/2$-concurrence $C_{1/2}\left(\rho_W\right)$ for $W\geq0.6$.

\section{summary}\label{summary}
We have introduced the concept of $\alpha$-concurrence and shown that the $\alpha$-concurrence is a well defined entanglement measure. Analytical lower bounds of the $\alpha$-concurrence for general mixed states have been derived based on PPT and realignment criterion. Specifically, we have derived explicit formulae for the $\alpha$-concurrence of isotropic states and Werner states. Interestingly our lower bounds are exact for isotropic states and Werner states with $d=2$. Our parameterized entanglement measure $\alpha$-concurrence gives a family of entanglement measures and enriches the theory of quantum entanglement, which may highlight further researches on the study of quantifying quantum entanglement and the related
investigations like monogamy and polygamy relations in entanglement distribution, as well as
the physical understanding of quantum correlations.

\bigskip
\section*{Acknowledgments}
This work is supported by the National Natural Science Foundation of China (NSFC) under Grant Nos. 12075159 and 12171044; Beijing Natural Science Foundation (Grant No. Z190005); Academy for Multidisciplinary Studies, Capital Normal University; Shenzhen Institute for Quantum Science and Engineering, Southern University of Science and Technology (No. SIQSE202001), the Academician Innovation Platform of Hainan Province.

\begin{appendix}
\section{Proof of Lemma 1}\label{a}
We only need to prove that $F_{\alpha}\left(t\rho+\left(1-t\right)\sigma\right)\geqslant t F_{\alpha}\left(\rho\right)+\left(1-t\right)F_{\alpha}\left(\sigma\right)$ for any $t\in\left[0,1\right]$. Since for any concave function $f$, the $\mathrm{Tr}\left[f\left(\rho\right)\right]$ is also concave \cite{RevModPhys.50.221},
$f\left(x\right)=x^{\alpha}$ with $x\in\left[0,1\right]$ is concave for any $0\leqslant\alpha\leqslant 1/2$. Let ${\rho}'=t\rho+\left(1-t\right)\sigma=\sum_j q_j|\phi_j\rangle\langle\phi_j|$, $\rho=\sum_ip_i|\psi_i\rangle\langle\psi_i|$ and $\sigma=\sum_k r_k|\xi_k\rangle\langle\xi_k|$ be corresponding eigendecompositions. We have
\begin{align}
F_{\alpha}\left({\rho}'\right)&=\sum_j\left[t\braket{\phi_j|\rho|\phi_j}
+\left(1-t\right)\braket{\phi_j|\sigma|\phi_j}\right]^{\alpha}-1\nonumber\\
&\geqslant \sum_j\left\{t\braket{\phi_j|\rho|\phi_j}^{\alpha}
+\left(1-t\right)\braket{\phi_j|\sigma|\phi_j}^{\alpha}\right\}-1\nonumber\\
&=t\sum_j\left(\sum_ip_i|\braket{\phi_j|\psi_i}|^2\right)^{\alpha}  \nonumber\\ &  \ \ \ +\left(1-t\right)\sum_j\left(\sum_kr_k|\braket{\phi_j|\xi_k}|^2\right)^{\alpha}-1\nonumber\\
&\geqslant t\sum_{i,j}|\braket{\phi_j|\psi_i}|^2p_i^{\alpha}
+\left(1-t\right)\sum_{j,k}|\braket{\phi_j|\xi_k}|^2r_k^{\alpha}-1\nonumber\\
&=t\sum_ip_i^{\alpha}+\left(1-t\right)\sum_kr_k^{\alpha}-1\nonumber\\
&=tF_{\alpha}\left(\rho\right)+\left(1-t\right)F_{\alpha}\left(\sigma\right),
\end{align}
where the two inequalities follow from the concavity of $f$. The equality holds if $\rho$ and $\sigma$ are identical. $\hfill\qedsymbol$

\section{$\eta_{\alpha}$ for Isotropic states}\label{b}
Let $\mathcal{T}_{iso}$ be the $\left(U\otimes U^*\right)$-twirling operator defined by $\mathcal{T}_{iso}\left(\rho\right)=\int dU\left(U\otimes U^{\ast}\right)\rho\left(U\otimes U^{\ast}\right)^{\dagger}$, where $dU$ denotes the standard Haar measure on the group of all $d\times d$ unitary operations. Then the operator satisfies that $\mathcal{T}_{iso}\left(\rho\right)=\rho_{F\left(\rho\right)}$ with $F\left(\rho\right)=\braket{\Psi|\rho|\Psi}$. One has $\mathcal{T}_{iso}\left(\rho_F\right)=\rho_F$ \cite{PhysRevA.64.062307,PhysRevA.59.4206,PhysRevA.68.062304}. Applying $\mathcal{T}_{iso}$ to the pure state $\ket{\psi}$ given in (\ref{ac1}),  $\ket{\psi}=\sum_{i=1}^r\sqrt{\lambda_i}U_A\otimes U_B\ket{ii}$ with $\ket{a_i}=U_A\ket{i}$ and $\ket{b_i}=U_B\ket{i}$, we have
\begin{equation}\label{p4iso1}
\mathcal{T}_{iso}\left(|\psi\rangle\langle\psi|\right)=\rho_{F\left(|\psi\rangle\langle\psi|\right)}=\rho_{F\left(\vec{\lambda},V\right)},
\end{equation}
where $V=U_A^{T}U_B$ and
\begin{equation}\label{p4iso2}
F\left(\vec{\lambda},V\right)=|\braket{\Psi|\psi}|^2=\frac{1}{d}\Big|\sum_{i=1}^r\sqrt{\lambda_i}V_{ii}\Big|^2
\end{equation}
with $V_{ij}=\braket{i|V|j}$, and  $\vec{\lambda}$ is the Schmidt vector of (\ref{ac1}). Then the function $\eta$ defined in (\ref{p30}) becomes
\begin{equation}\label{p4iso3}
\eta_{\alpha}\left(\rho_F\right)=\min_{\set{\vec{\lambda},V}}\left\{C_{\alpha}\left(\vec{\lambda}\right):\frac{1}{d}\Big|\sum_{i=1}^r\sqrt{\lambda_i}V_{ii}\Big|^2=F\right\}.
\end{equation}
It has been proved that the minimum above is attained for $V=I$ \cite{PhysRevA.68.062304}. Therefore, we have
\begin{align}\label{p4iso8}
\eta_{\alpha}\left(\rho_F\right)=\min_{\vec{\lambda}}\left\{C_{\alpha}\left(\vec{\lambda}\right):\frac{1}{d}\Big|\sum_{i=1}^r\sqrt{\lambda_i}\Big|^2=F\right\}.
\end{align}

For $F\in(0,\frac{1}{d}]$, one can always chose suitable $U_A$ and $U_B$ such that $\lambda_1=1$, and hence $\eta_{\alpha}\left(\rho_F\right)=0$. For $F\in(\frac{1}{d},1]$, similar to \cite{PhysRevA.103.052423,PhysRevLett.85.2625}, by using the Lagrange multipliers \cite{PhysRevA.67.012307} one can minimize (\ref{p4iso8}) subject to the constraints
\begin{align}
\sum_{i=1}^r\lambda_i=1,~~~
\sum_{i=1}^r\sqrt{\lambda_i}=\sqrt{Fd}
\end{align}
with  $Fd\geqslant 1$. An extremum is attained when
\begin{equation}\label{B6}
\left(\sqrt{\lambda_i}\right)^{2\alpha-1}+\Lambda_1\sqrt{\lambda_i}+\Lambda_2=0,
\end{equation}
where $\Lambda_1$ and $\Lambda_2$ denote the Lagrange multipliers.

It is evident that $f\left(\sqrt{\lambda_i}\right)=\left(\sqrt{\lambda_i}\right)^{2\alpha-1}$ is a convex function of $\sqrt{\lambda_i}$ for any $0\leqslant \alpha\leqslant 1/2$. Since a convex function and a linear function cross at most two points, equation (\ref{B6}) has at most two possible nonzero solutions for $\sqrt{\lambda_i}$. Let $\gamma$ and $\delta$ be these two positive solutions with $\gamma>\delta$. The Schmidt vector $\vec{\lambda}=\left \{ \lambda_1, \lambda_2,\dots,\lambda_r,0,\cdots,0 \right \} $ has the form,
\begin{equation}
\lambda_j=\begin{cases}
\gamma^2,&j=1,2,...,n, \\
\delta^2, &j=n+1,...,n+m, \\
 0,  &j=n+m+1,...,d,
\end{cases}
\end{equation}
where $r=n+m\leqslant d$ and $n\geqslant 1$. The minimization problem of (\ref{p4iso8}) has been reduced to the following minimum problem,
\begin{equation}\label{app2}
\eta_{\alpha}\left(\rho_F\right)=\min_{n,m}C_{\alpha}^{nm}\left(F\right)
\end{equation}
with
\begin{equation}\label{b7}
C_{\alpha}^{nm}\left(F\right)=n\gamma^{2\alpha}+m\delta^{2\alpha}-1,
\end{equation}
subject to the constraints
\begin{align}\label{b8}
n\gamma^2+m\delta^2=1,~~~
n\gamma+m\delta=\sqrt{Fd}.
\end{align}
By solving Eq. (\ref{b8}), we obtain
\begin{equation}\label{b9}
\gamma_{nm}^{\pm}\left(F\right)=\frac{n\sqrt{Fd}
\pm\sqrt{nm\left(n+m-Fd\right)}}{n\left(n+m\right)},
\end{equation}
\begin{equation}\label{b10}
\delta_{nm}^{\pm}\left(F\right)=\frac{m\sqrt{Fd}
\mp\sqrt{nm\left(n+m-Fd\right)}}{m\left(n+m\right)}.
\end{equation}

The relation $\gamma_{nm}^{\pm}\left(F\right)=\delta_{mn}^{\mp}\left(F\right)$ suggests that we only need to consider the cases $\gamma_{nm}:=\gamma_{nm}^+\left(F\right)$ and $\delta_{nm}=\delta_{nm}^+\left(F\right)$, which are real for $Fd\leqslant n+m$. On the other hand, since $\delta_{nm}$ should be non-negative, we must have $Fd\geqslant n$. Therefore, we see that $\delta_{nm}\leqslant \sqrt{Fd}/\left(n+m\right)\leqslant \gamma_{nm}$, in consistent with the assumption $\gamma>\delta$. Here, $n\geqslant 1$ as $n=0$ is ill defined.

We seek is the minimum of $C_{\alpha}^{nm}\left(F\right)$ over all possible $n$ and $m$, by  minimizing $C_{\alpha}^{nm}$ on the parallelogram defined by $1\leqslant n\leqslant Fd$ and $Fd\leqslant n+m\leqslant d$. Note that the parallelogram collapses to a line when $Fd = 1$, i.e., the separable boundary. We have $\gamma_{nm}\geqslant\delta_{nm}\geqslant 0$ in the  parallelogram. Moreover, $\gamma_{nm}=\delta_{nm}$ if and only if $n+m=Fd$; while $\delta_{nm}=0$ if and only if $n=Fd$.

When $\alpha=1/2$, we see from Eqs. (\ref{b7}) and (\ref{b8}) that Eq. (\ref{text32}) holds without any optimization. When $\alpha=0$, $C_0^{nm}\left(F\right)=n+m-1$ and Eq. (\ref{text32}) satisfied with the constraint conditions. From Eq.(\ref{b8}) the derivatives of $\gamma_{nm}$ and $\delta_{nm}$ with respect to $n$ and $m$ are given by,
\begin{align}\label{b11}
\frac{\partial\gamma}{\partial n}&=\frac{1}{2n}\frac{2\gamma\delta-\gamma^2}{\gamma-\delta},\nonumber\\
\frac{\partial\delta}{\partial n}&=-\frac{1}{2m}\frac{\gamma^2}{\gamma-\delta},\nonumber\\
\frac{\partial\delta}{\partial m}&=-\frac{1}{2m}\frac{2\gamma\delta-\delta^2}{\gamma-\delta},\nonumber\\
\frac{\partial\gamma}{\partial m}&=\frac{1}{2n}\frac{\delta^2}{\gamma-\delta}.
\end{align}
Hence, using Eq. (\ref{b7}) we have the partial derivatives of $C_{\alpha}^{nm}\left(F\right)$ with respect to $n$ and $m$,
\begin{equation}\label{b12}
\frac{\partial C_{\alpha}^{nm}}{\partial n}=\left(1-\alpha\right)\gamma^{2\alpha}+\alpha\gamma^2\delta\frac{\gamma^{2\alpha-2}-\delta^{2\alpha-2}}{\gamma-\delta},
\end{equation}
\begin{equation}\label{b13}
\frac{\partial C_{\alpha}^{nm}}{\partial m}=\left(1-\alpha\right)\delta^{2\alpha}+\alpha\delta^2\gamma\frac{\gamma^{2\alpha-2}-\delta^{2\alpha-2}}{\gamma-\delta}.
\end{equation}

By lengthy calculations, we have
\begin{equation}\label{b14}
\frac{\partial C_{\alpha}^{nm}}{\partial m}=\frac{\delta^{2\alpha+1}}{\gamma-\delta}\left\{\left(1-\alpha\right)\left(\frac{\gamma}{\delta}-1\right)+\alpha\left(\frac{\gamma}{\delta}\right)^{2\alpha-1}-\alpha\frac{\gamma}{\delta}\right\}.
\end{equation}
Denote $t=\frac{\gamma}{\delta}$. One has $t\geqslant 1$. Let
\begin{equation}\label{b15}
g\left(t\right)=\left(1-\alpha\right)\left(t-1\right)+\alpha t^{2\alpha-1}-\alpha t.
\end{equation}
We have $g\left(1\right)=0$ and
\begin{equation}\label{b16}
\frac{\partial g}{\partial t}=\left(1-2\alpha\right)\left(1-\alpha t^{2\alpha-2}\right).
\end{equation}
Set $h\left(t\right)=1-\alpha t^{2\alpha-2}$ with $h\left(1\right)=1-\alpha> 0$. We obtain
\begin{equation}\label{b17}
\frac{\partial h}{\partial t}=-\alpha\left(2\alpha-2\right)t^{2\alpha-3}\geqslant 0.
\end{equation}
From Eqs. (\ref{b17})  and  (\ref{b16}), combining with Eq.(\ref{b15}) we have
\begin{equation}\label{b18}
\frac{\partial C_{\alpha}^{nm}}{\partial m}\geqslant 0.
\end{equation}

Now corresponding to moving perpendicularly to and parallel to the $n+m=$constant boundaries of the parallelogram, we make a parameter transformation, $u=n-m$ and $v=n+m$. The derivative of $C_{\alpha}^{nm}$ with respect to $u$ is given by
\begin{align*}
&\frac{\partial C_{\alpha}^{nm}}{\partial u}=\frac{1}{2}\left(\frac{\partial C_{\alpha}}{\partial n}-\frac{\partial C_{\alpha}}{\partial m}\right)\\
&=\frac{1}{2}\left\{\gamma^{2\alpha-1}\left[\left(1-\alpha\right)\gamma+\alpha\delta\right]-\delta^{2\alpha-1}\left[\left(1-\alpha\right)\delta+\alpha\gamma\right]\right\}\\
&=\frac{\gamma^{2\alpha}}{2}\left\{1-\alpha+\alpha\frac{\delta}{\gamma}-\left(\frac{\delta}{\gamma}\right)^{2\alpha}\left(1-\alpha+\alpha\frac{\gamma}{\delta}\right)\right\}.
\end{align*}

Set $x=\frac{\delta}{\gamma}$ with $0\leqslant x\leqslant 1$. Let
\begin{equation}\label{b19}
f\left(x\right)=1-\alpha+\alpha x - x^{2\alpha}\left(1-\alpha+\alpha x^{-1}\right)
\end{equation}
with $f\left(1\right)=0$. We have the derivative of $f\left(x\right)$ respect to $x$,
\begin{equation}\label{b20}
\frac{\partial f}{\partial x}=\alpha\left\{1-\left[2x^{2\alpha-1}\left(1-\alpha+\alpha x^{-1}\right)-x^{2\alpha-2}\right]\right\}.
\end{equation}
Again let
\begin{equation}\label{b21}
k\left(x\right)=2x^{2\alpha-1}\left(1-\alpha+\alpha x^{-1}\right)-x^{2\alpha-2}
\end{equation}
with $k\left(1\right)=1$. Then
\begin{equation}\label{b22}
\frac{\partial k}{\partial x}=x^{2\alpha-3}l\left(x\right),
\end{equation}
where
\begin{equation}\label{23}
l\left(x\right)=2\left(2\alpha-1\right)\left(x-\alpha x\right)-\left(1-\alpha\right)\left(4\alpha-2\right).
\end{equation}
We have $l\left(1\right)=0$ and
\begin{equation}\label{b24}
\frac{\partial l}{\partial x}=2\left(2\alpha-1\right)\left(1-\alpha\right)\leqslant 0.
\end{equation}

From Eqs. (\ref{b24}) and (\ref{b22}), combining Eq. (\ref{b20}) we obtain $\partial f/\partial x \geqslant 0$. Then for any $x\in [0,1]$ we have $f\left(x\right)\leqslant f\left(1\right)=0$. Therefore,
\begin{equation}\label{b25}
\frac{\partial C_{\alpha}^{nm}}{\partial u}\leqslant 0.
\end{equation}
From Eqs. (\ref{b18}) and (\ref{b25}), the minimum of $C_{\alpha}^{nm}\left(F\right)$ is obtained when $m$ is the minimum and $u$ is the maximum. These results imply that the minimum of  $C_{\alpha}^{nm}\left(F\right)$ occurs at the vertex of $n=Fd$ and $m=0$. Specifically, since $\gamma_{nm}=\delta_{nm}$ on the boundary $n+m=Fd$ where Eqs. (\ref{b18}) and (\ref{b25}) are both hold, we have ${n}'=n+m=Fd$ and ${m}'=0$. In this way, we derive an analytical expression of the function $\eta_{\alpha}\left(\rho_F\right)$ as follows,
\begin{equation}\label{b27}
\eta_{\alpha}\left(\rho_F\right)=\left(Fd\right)^{1-\alpha}-1.
\end{equation}

\section{$\eta_{\alpha}$ for Werner states}\label{c}
Let $\mathcal{T}_{wer}\left(\rho\right)=\int dU\left(U\otimes U\right)\rho\left(U^{\dagger}\otimes U^{\dagger}\right)$ be the $\left(U\otimes U\right)$-twirling transformations \cite{PhysRevA.64.062307}. Then the Werner states defined in (\ref{p4wer1}) satisfy that, in analogous to the isotropic states, $\mathcal{T}_{wer}\left(\rho\right)=\rho_{W\left(\rho\right)}$, where $W\left(\rho\right)=\mathrm{Tr}\left(\rho\sum_{i<j}|\Psi_{ij}^-\rangle\langle\Psi_{ij}^-|\right)$ and $\mathcal{T}_{wer}\left(\rho_W\right)=\rho_W$ \cite{PhysRevA.59.4206,PhysRevA.68.062304}. Applying $\mathcal{T}_{wer}$ to the pure state $\ket{\psi}$ defined in (\ref{ac1}), $\ket{\psi}=\sum_{i=1}^r\sqrt{\lambda_i}U_A\otimes U_B\ket{ii}$, we have
\begin{equation}\label{ap3w1}
\mathcal{T}_{wer}\left(|\psi\rangle\langle\psi|\right)=\rho_{W\left(|\psi\rangle\langle\psi|\right)}=\rho_{W\left(\vec{\lambda},\Lambda\right)},
\end{equation}
where $\Lambda=U_A^{\dagger}U_B$ and
\begin{align}\label{ap3w2}
W\left(\vec{\lambda},\Lambda\right)&=\sum_{i<j}|\braket{\Psi_{ij}^-|\psi}|^2\nonumber\\
&=\frac{1}{2}\sum_{i<j}|\sqrt{\lambda_i}\Lambda_{ji}-\sqrt{\lambda_j}\Lambda_{ij}|^2
\end{align}
with $\Lambda_{ij}=\braket{i|\Lambda|j}$. Then the function $\eta$ defined in (\ref{p30}) becomes
\begin{equation}\label{ap3w3}
\eta_{\alpha}\left(\rho_W\right)=\min_{\set{\vec{\lambda},\Lambda}}\left\{C_{\alpha}\left(\vec{\lambda}\right):W\left(\vec{\lambda},\Lambda\right)=W\right\}.
\end{equation}
By $W\left(\vec{\lambda},\Lambda\right)=W$ we have
\begin{align}\label{ap3w4}
2W&=1-\sum_{i=1}^r\lambda_i|\Lambda_{ii}|^2-2\sum_{i<j}\sqrt{\lambda_i\lambda_j}\mathrm{Re}\left(\Lambda_{ij}\Lambda_{ji}^{\ast}\right)\nonumber\\
&\leqslant1+2\sum_{i<j}\sqrt{\lambda_i\lambda_j}|\mathrm{Re}\left(\Lambda_{ij}\Lambda_{ji}^{\ast}\right)|\nonumber\\
&\leqslant1+2\sum_{i<j}\sqrt{\lambda_i\lambda_j}\nonumber\\
&=|\sum_{i=1}^r\sqrt{\lambda_i}|^2,
\end{align}
where $\mathrm{Re}\left(z\right)$ is the real part of $z$.

Note that the equalities in (\ref{ap3w4}) hold if only the two nozero components $\Lambda_{01}=1$ and $\Lambda_{10}=-1$, and $\vec{\lambda}=\left(\lambda_1,\lambda_2,0,\cdots,0\right)$, which give rise to the optimal minimum of (\ref{ap3w3}) \cite{PhysRevA.64.062307}. Therefore, (\ref{ap3w3}) becomes
\begin{equation}\label{ap3w5}
\eta_{\alpha}\left(\rho_W\right)=\min_{\vec{\lambda}}\left\{C_{\alpha}\left(\vec{\lambda}\right):|\sum_{i=1}^2\sqrt{\lambda_i}|^2=2W\right\}.
\end{equation}
For $W\in(0,\frac{1}{2}]$, we can always chose suitable $U_A$ and $U_B$ to have that  $\lambda_1=1$, which results in $\eta_{\alpha}\left(\rho_W\right)=0$. For $W>1/2$, one minimizes (\ref{ap3w5}) subject to the constraints
\begin{align}
\sum_{i=1}^2\lambda_i=1,~~~\sum_{i=1}^2\sqrt{\lambda_i}=\sqrt{2W}
\end{align}
with $W>1/2$.
The rest of the calculation is the similar to the one in Appendix \ref{b}. We only need to set $d=2$ and $F=W$. In this way, we can obtain
\begin{equation}\label{ap3w6}
\eta_{\alpha}\left(\rho_W\right)=\left(2W\right)^{1-\alpha}-1.
\end{equation}

\end{appendix}

\end{document}